\documentclass[conference]{IEEEtran}

\usepackage{hyperref}
\usepackage{url}
\usepackage{amsmath,amssymb,amsfonts,amsthm,amsbsy,bm}
\usepackage{epsfig}
\usepackage{cite}
\usepackage{microtype}
\usepackage{algorithm}
\usepackage[noend]{algorithmic}
\newcommand{\be}{\begin{equation}}
\newcommand{\ee}{\end{equation}}

\ifCLASSINFOpdf
\else
\fi

\begin{document}
\title{Bayesian spike inference from calcium imaging data}

\author{\IEEEauthorblockN{Eftychios A. Pnevmatikakis, Josh Merel, Ari Pakman and Liam Paninski}
\IEEEauthorblockA{Department of Statistics\\
Center for Theoretical Neuroscience \\
Grossman Center for the Statistics of Mind\\
Columbia University, New York, NY}}

\maketitle

\begin{abstract}
%\bold math
We present efficient Bayesian methods for extracting neuronal spiking information from calcium imaging data. The goal of our methods is to sample from the posterior distribution of spike trains and model parameters (baseline concentration, spike amplitude etc) given noisy calcium imaging data. We present discrete time algorithms where we sample the existence of a spike at each time bin using Gibbs methods, as well as continuous time algorithms where we sample over the number of spikes and their locations at an arbitrary resolution using Metropolis-Hastings methods for point processes. We provide Rao-Blackwellized extensions that (i) marginalize over several model parameters and (ii) provide smooth estimates of the marginal spike posterior distribution in continuous time. Our methods serve as complements to standard point estimates and allow for quantification of uncertainty in estimating the underlying spike train and model parameters.
\end{abstract}

%\IEEEpeerreviewmaketitle

\section{Introduction}
Calcium imaging is an increasingly popular technique for large scale data acquisition in neuroscience \cite{AHR13}.  The method detects underlying, single neuron activity indirectly through observations of fluorescent indicators for calcium concentration. A key problem in the analysis of calcium imaging data is the inference of exact spike times from the noisy calcium signal which has slower dynamics compared to neural spiking and is sampled at a relatively low acquisition rate. A variety of methods have been proposed to deal with this problem including particle filtering \cite{vogelstein2009spike}, fast nonnegative deconvolution for approximate maximum-a-posteriori (MAP) inference \cite{Vogelstein10}, greedy template matching \cite{grewe2010}, and methods for estimating signals with finite rate of innovation \cite{OSD13}. In these methods, parameter estimation is typically performed offline or in an iterative manner using for example the expectation maximization algorithm.

In this paper we propose Bayesian methods for sampling from the joint posterior distribution of the spike train and the model parameters given the fluorescence observations. We present two efficient approaches for sampling the spikes. The first is a discrete time binary sampler that samples whether a spike occurred at each timebin using Gibbs sampling. %either Gibbs sampling or Hamiltonian Monte Carlo (HMC). 
By exploiting the weak interaction between spikes at distant timebins, we show that a full sample can be obtained with just $O(T)$ complexity, where $T$ is the number of timebins, and that parallelization is also possible. Our second sampler operates in continuous time and samples the number of spikes and the spike times at arbitrary resolution using Metropolis-Hastings (MH) techniques. We use a proposal distribution to move the spike times around a local neighborhood that is based on the resulting signal residual. This proposal distribution enables fast mixing and tractable inference; each full sample is obtained with just $O(K)$ complexity where $K$ is the total number of spikes, rendering this algorithm particularly efficient for recordings that are sparse and/or are obtained at a fine resolution. Moreover, in high-SNR conditions, this method enables super-resolution spike inference (i.e. determining where each spike occurred within each timebin) and smooth estimates of the marginal spike posterior using a Rao-Blackwellized scheme. We also show that is possible to marginalize over several of the model parameters and derive collapsed Gibbs samplers that exhibit faster mixing.% rates.

\section{Model description and block Gibbs sampling}
\label{model}
We assume that we observe a single neuron calcium trace for a duration of $T$ timesteps. Let $\bm{s}\in\{0,1\}^T$ the binary spiking vector of the neuron that indicates the existence of the spike at each timebin. The calcium activity that is generated by $\bm{s}$ can be described by a simple first order autoregressive process as
\[%be
	c(t) = \gamma c(t-1) + As(t),
\]%ee
where $\gamma$ is a discrete time constant with $0<\gamma<1$, $A$ indicates the amplitude of each spike, and $c(1) = c_1 + As(1)$, with $c_1$ an initial condition for the calcium concentration. Our fluorescence observation vector $\bm{y}$, can be written as 
\[%be
	y(t) = c(t) + b + \varepsilon_t,
\]%ee
where $b$ is the baseline concentration and $\varepsilon_t \sim\mathcal{N}(0,\sigma^2)$ is some random Gaussian noise. Our goal is to estimate the spiking vector $\bm{s}$ given the observation vector $\bm{y}$, which we assume is normalized in the interval $[0,1]$. Approximate MAP methods \cite{Vogelstein10} have been shown to perform well under high SNR assumptions. However in the low SNR regime their performance degrades, and the parameter estimation becomes more challenging. To overcome these limitations we introduce a block-Gibbs sampler that produces samples from the joint posterior distribution of the spikes and model parameters given $\bm{y}$. Let $G\in\mathbb{R}^{T\times T}$ and $v\in\mathbb{R}^T$ defined respectively as 
\[%be
	G = \left[\begin{array}{cccc} 1 & 0 & \ldots & 0 \\ -\gamma & 1  & \ldots & 0  \\ \vdots & \ddots & \ddots & \vdots \\ 0 & \ldots & -\gamma & 1\end{array}\right], \quad \bm{v} = \left[\begin{array}{c} 1 \\ \gamma \\ \vdots \\ \gamma^{T-1}\end{array}\right].
\]
By denoting $\bm{\theta} = [A,b,c_1]^T$, the likelihood can be written as
\[
	p(\bm{y}|\bm{s},\bm{\theta},\sigma^2) \propto \exp\left(-\frac{A^2}{2\sigma^2}\bm{s}^TG^{-T}G^{-1}\bm{s} + \frac{A}{\sigma^2}\bm{s}^TG^{-T}\bm{\tilde{y}}\right),
\]
with $\bm{\tilde{y}} = \bm{y} - b\bm{1}_T - c_1\bm{v}$ ($\bm{1}_T$ denotes a vector of ones of length T).
We here place an i.i.d. Bernoulli process prior on the spike trains so the probability of a spike in any timebin is $\pi$. Under this uniform spiking assumption the discrete time constant $\gamma$ can be estimated robustly from the autocovariance function of $\bm{y}$. For the prior probability $\pi$ we set a hyper-prior $\pi\sim \mathrm{Beta}(\alpha,\beta)$. At each iteration we update the parameters $\alpha,\beta$ using empirical Bayes \cite{casella2001}:  For a spiking vector $\bm{s}$ the evidence function \cite{Bishop06} can be written as
\[ 
%\begin{split}
%	p(\bm{s}|\alpha,&\beta) = \int_{0}^{1}p(\bm{s}|\pi)p(\pi|\alpha,\beta)d\pi \\
%	& = \frac{\Gamma(\alpha+\beta)}{\Gamma(\alpha)\Gamma(\beta)}\int_{0}^{1}\pi^{\bm{1}^T_T\bm{s}+\alpha-1}(1-\pi)^{T - \bm{1}^T_T\bm{s} + \beta -1}d\pi \\
%	& = \dfrac{\alpha^{\bm{1}^T_T\bm{s}}\beta^{T - \bm{1}^T_T\bm{s}}}{(\alpha+\beta)^T} = \dfrac{(\alpha/\beta)^{\bm{1}^T_T\bm{s}}}{(1+\alpha/\beta)^T}.
%\end{split} %\label{evidence} 
	p(\bm{s}|\alpha,\beta) = \int_{0}^{1}p(\bm{s}|\pi)p(\pi|\alpha,\beta)d\pi = \dfrac{(\alpha/\beta)^{\bm{1}^T_T\bm{s}}}{(1+\alpha/\beta)^T}.
\]
The evidence function is constant for a fixed ratio $r = \alpha/\beta$ and is maximized for $r = \bm{1}^T_T\bm{s}/(T-\bm{1}^T_T\bm{s})$. To find distinct values for $\alpha$ and $\beta$ we place a flat hyperprior on $\beta$, which yields an exponential posterior $\beta | \pi,r \sim \mathrm{Exp}(-\log(\pi)r -\log(1-\pi))$. 
For the rest of parameters $\bm{\theta}$ we assume a joint half-normal (nonnegative) distribution, $\bm{\theta} \sim \mathcal{N}(\bm{\mu},\Sigma)1_{\{\bm{\theta}\in\mathbb{R}^3_+\}}$. The parameters $\bm{\mu}$ and $\Sigma$ can also be learned from the data, but we selected general values that assume little prior knowledge and model a wide prior. Other prior choices, e.g. exponential, yield practically the same results. Finally, for the noise variance $\sigma^2$ we set an inverse Gamma prior $\sigma^2\sim \mathrm{InvGamma}(1,0.1)$\protect\footnote{The variance $\sigma^2$ can also be estimated from the autocovariance function via the Yulle-Walker equations but we favor a Bayesian approach here.}, which is a weak and relatively flat prior for both high and low-SNR regions for traces $\bm{y}$ normalized to the $[0,1]$ interval. Under these assumptions, the block Gibbs sampler proceeds as follows to draw samples from the joint posterior:
\[	\begin{split}
	\log p(\bm{s}|\pi,\bm{\theta},&\sigma^2,\bm{y})  \propto   -\frac{1}{2}\bm{s}^TW\bm{s} + \bm{s}^T\left( \frac{A}{\sigma^2}G^{-T}\bm{\tilde{y}} + l\bm{1}_T\right) \\
	r & =  \bm{1}^T_T\bm{s}/(T-\bm{1}^T_T\bm{s})\\ 
	\beta | \pi,r & \sim \mathrm{Exp}(-\log(\pi)r -\log(1-\pi)) \\
	\alpha & = r\beta \\
	\pi|\bm{s}  & \sim \mathrm{Beta}(\bm{1}^T\bm{s} + \alpha, T - \bm{1}^T\bm{s} + \beta) \\
	\bm{\theta}|\bm{s},\sigma^2,\bm{y}  & \sim \mathcal{N}(\Lambda(\sigma^{-2}S^T\bm{y} +\Sigma^{-1}\bm{\mu}),\Lambda)1_{\{\bm{\theta}\in\mathbb{R}^3_+\}} \\
	\sigma^2 | \bm{s},\bm{\theta},\bm{y} & \sim \mathrm{InvGamma}(1+T/2,0.1 + \|\bm{y} - S\bm{\theta}\|^2/2),
\end{split} \]
with $W = A^2(G^{-T}G^{-1})/\sigma^2$, $l = \log (\pi/(1-\pi))$, $S$ is a $T\times3$ matrix that depends on the current state of $\bm{s}$, given by
\[
	S = \left[ G^{-1}\bm{s}, \bm{1}_T, \bm{v}\right],
\]
%\be \begin{split}
%	S(t,1) = \gamma S(t-1,1) + s(t), \quad S(t,2) = 1, \quad S(t,3) = \gamma^{t-1},
%\end{split} \ee
and $\Lambda = (\Sigma^{-1}+\sigma^{-2}S^TS)^{-1}$. We now turn to the main problem of sampling from the posterior of the spike vector $\bm{s}$.

\section{Discrete binary sampler}
\label{discrete}

%\subsection{Metropolized-Gibbs sampler}
To sample $\bm{s}$ we take a Gibbs-sampling approach where at each iteration we sample $s(i)$ given the current state of the rest of the entries. % $\bm{s}_{\backslash i}$. 
We use a Metropolized Gibbs (MG) sampler, with the flip of each binary entry being the possible move \cite{liu96}. If $s_i^{(k)}$ is the state of $s(i)$ at the $k$-th sample, and define $\bm{s}_{\mathrm{cur}} = [\bm{s}_{1:i-1}^{(k)},\bm{s}_{i:T}^{(k-1)}]^T$ the current state of $\bm{s}$. The sampling algorithm flips $s_i^{(k)}$ with probability
\be
	%\mathbb{P}(s_i^{(k)} = 1 - s_i^{(k-1)}) =	 
	\mathbb{P}_{\text{flip}}(i) = \min\left(1, \frac{\mathbb{P}(\bm{s}_{\mathrm{flipped}})}{\mathbb{P}(\bm{s}_{\mathrm{cur}})}\right)
	\label{ratio}
\ee
The ratio inside \eqref{ratio} can be more efficiently computed in the log-domain. By denoting $\bm{\alpha} =   \frac{A}{\sigma^2}G^{-T}\bm{\tilde{y}} + l\bm{1}_T$ the log-ratio becomes
%\be
%	\mathbb{P}_{\text{flip}}(i) = \min(1,\exp((1-2s_i^{(k-1)})(\alpha(i) - A^2W(\cdot,i)\bm{\tilde{s}}))),
%\ee
\be
	(1-2s_i^{(k-1)})((2s_i^{(k-1)}-1)[W]_{ii}/2 - \bm{s_{\mathrm{cur}}}^TW\bm{e_i} + a(i)),
	\label{log-ratio}
\ee
with $\bm{e_i}$ the $i$-th standard basis vector. It is important to note that the log-ratio of \eqref{log-ratio} can be computed efficiently in just $O(1)$ time. The matrix $G^{-1}$ is lower triangular and Toeplitz with $[G^{-1}]_{ij} = \gamma^{i-j}1_{i\geq j}$. Therefore it can be approximated by a banded Toeplitz matrix with bandwidth that depends on $\gamma$. Consequently $W$ can again be approximated by a banded matrix and is also asymptotically Toeplitz for $T\rightarrow \infty$. It follows that the products $\bm{s_{\mathrm{cur}}}^TW\bm{e_i}$ can be computed in just $O(1)$ time (technically in $O(m/\log(1/\gamma))$ for $m$ bits of accuracy). This gives a total $O(T)$ complexity per full sweep to obtain a new sample from $p(\bm{s}|\bm{y},p,\bm{\theta},\sigma^2)$. For large $T$, the algorithm can also be parallelized -- notice that the entries of $\bm{s}$ which are at least $L = O(1/\log(1/\gamma))$ timesteps apart, are approximately conditionally independent. $\bm{s}$ can be partitioned into chunks of length $L$, that can be sampled in parallel. It is obvious that instead of using the MG sampler presented here, one can also use a plain Gibbs sampler with similar $O(T)$ complexity. In practice we observed that the MG sampler led to faster mixing.% rates.

%\subsection{Augmented Gibbs Sampler}
%An alternative  \cite{MS10}

%\subsection{Exact Hamiltonian Monte Carlo sampler}
%An alternative to Gibbs sampling is HMC that can often lead to faster mixing rates. For the case of binary distributions, exact HMC is possible using the continuous augmentation technique presented in \cite{Ari13}. Applied to our model, HMC yields qualitatively similar results to the Gibbs sampler (not shown).

\section{Continuous time sampler}
The spike samplers presented in section \ref{discrete} operate on a discrete domain, with the length of each timebin set by the frame rate of the calcium imaging. While simple and effective, these methods can have several disadvantages. First, when the calcium signal is acquired by raster scanning, the frame rate is typically in the range of 10-30Hz. The length of each timebin is too large to assume that the underlying neuron can fire at most 1 spike per bin. In certain applications, higher resolution of the spike time is needed. A typical example is in ``connectomics" where the order with which the various neurons fire spikes is crucial for determining their connectivity \cite{Mishchenko2011,KPP13}. Moreover, the complexity of the discrete samplers scales with the number of observed timebins (i.e. with the temporal resolution). In experiments where this resolution is very fine discrete, binary samplers can become computationally expensive. Compounding this, neural activity may be very sparse and thus sampling at intervals where no spikes occur is uninformative. Instead, a method that scales with the number of spikes is more desirable.

To address these issues we propose a sampler that samples directly the spike times in continuous time, given the calcium observations. The state space corresponds to the set of spike times, and transdimensional moves change the dimensionality of this state space.  At every iteration we sample the new time of the $i$-th spike $t_i$ given the current location of the rest of the spikes and the hyperparameters. Since the number of spikes $K$ is in general unknown, we also allow for spikes insertions and deletions at each iteration. Similar approaches have been proposed before, see e.g. \cite{TG08} for the context of signals with finite rate of innovation, where however the number of ``spikes" $K$ was considered known.
In the continuous time setup, the calcium evolution is described by the differential equation
\[
	\dfrac{d}{dt}c(t) = -\dfrac{1}{\tau}c(t) + As(t),
\]
where $\tau$ is the (continuous) time constant of the calcium indicator. For a timebin of size $\Delta$ with $\Delta \ll \tau$ the discrete and continuous time constants are related through $\gamma = 1 - \Delta/\tau$. If $t_1,\ldots,t_k$ are the spike times, denoted by $\mathfrak{T}$, the calcium signal is given by
\be
	c_{\mathfrak{T}}(t) = c_1e^{-\frac{t}{\tau}} + A\sum_{k=1}^Te^{-\frac{t-t_k}{\tau}}1_{\{t\geq t_k\}},
	\label{cal-eq}
\ee
where the subscript $\mathfrak{T}$ indicates the dependance of the calcium signal on the set of spike times $\mathfrak{T}$, and the observations are interpreted as $y(n) = c(n\Delta) + b + \varepsilon_n$, giving a likelihood of the form (we ignore $c_1$ for simplicity)
\[
	\log p(\bm{y}|\mathfrak{T},\bm{\theta},\sigma^2) \propto - \frac{1}{2\sigma^2}\|\bm{y} - b\bm{1}_T - \bm{c}_{\mathfrak{T}}\|^2,%\sum_{n=1}^T(y(n) - b - c(n\Delta))^2.
\]%ee
where $\bm{c}_{\mathfrak{T}} = [c_{\mathfrak{T}}(\Delta),\ldots,c_{\mathfrak{T}}(T\Delta)]^T$.
Finally, since the algorithm now operates in the continuous domain, we replace the  i.i.d. Bernoulli prior for the spikes with a homogeneous continuous Poisson process prior with parameter $\lambda$. If we set a prior $\lambda \sim \mathrm{Gamma}(\alpha,\beta)$ (with $\beta$ denoting the rate), then 
\[ \lambda | \mathfrak{T} \sim \mathrm{Gamma}(\alpha+|\mathfrak{T}|,\beta+T\Delta).\] 
For the hyper-parameters $\alpha,\beta$ we can compute again the evidence function:
\[ 
%\begin{split}
%	p(\mathfrak{T}|\alpha,\beta) &= \int_0^{\infty}p(\mathfrak{T}|\lambda)p(\lambda|\alpha,\beta)d\,\lambda \\
%				  & = \dfrac{\beta^{\alpha}}{\Gamma(\alpha)}\int_0^{\infty}\lambda^{|\mathfrak{T}|}e^{- \lambda T\Delta}\lambda^{\alpha-1}e^{-\beta\lambda}d\,\lambda \\
%				  & = \dfrac{\beta^{\alpha}\Gamma(\alpha+|\mathfrak{T}|)}{(\beta+T\Delta)^{a+|\mathfrak{T}|}\Gamma(\alpha)}
%\end{split} 
	p(\mathfrak{T}|\alpha,\beta) = \int_0^{\infty}p(\mathfrak{T}|\lambda)p(\lambda|\alpha,\beta)d\lambda = \dfrac{\beta^{\alpha}\Gamma(\alpha+|\mathfrak{T}|)}{(\beta+T\Delta)^{a+|\mathfrak{T}|}\Gamma(\alpha)}.
\]
$p(\mathfrak{T}|\alpha,\beta)$ is maximized for $\alpha,\beta \rightarrow \infty$ with $\beta = \alpha\Delta T/|\mathfrak{T}|$. By choosing these values the posterior of $\lambda$ puts all the mass in the maximum likelihood estimate $\lambda_{\mathrm{MLE}} = |\mathfrak{T}|/\Delta T$. We can either use this deterministic value or assume a fixed finite value for $\alpha$ and set $\beta = \alpha\Delta T/|\mathfrak{T}|$ at each iteration. We next describe the sampling of the spike times in more detail.

\subsection{Move existing spikes}
We can update individual spike times either using random walk Metropolis-Hastings (MH) moves or a Gibbs algorithm. For the MH algorithm, we propose a move $t_i \rightarrow t_i'$, using a Gaussian density, centered at $t_i$ and with standard deviation equal to e.g. $10\Delta$, as a proposal distribution. Since this proposal is symmetric, the proposed move is accepted with probability
\be
	\mathbb{P}(t_i\rightarrow t_i') = \min\left(1,\frac{p(\mathfrak{T}_{\mathrm{new}}|\bm{y},\bm{\theta},\sigma^2)}{p(\mathfrak{T}_{\mathrm{cur}}|\bm{y},\bm{\theta},\sigma^2)}\right).
	\label{MHratio}
\ee
Similarly to the  discrete case, each spike contributes an exponentially decaying calcium trace \eqref{cal-eq}, that has a temporally localized effect to the whole vector $\bm{c}_{\mathfrak{T}}$. By exploiting this, we can implement the spike shifting operation in just $O(1)$ time, provided that we keep in memory the residual vector $\bm{w}\in\mathbb{R}^T$ with $\bm{w} = \bm{y} - b\bm{1}_T - \bm{c}_{\mathfrak{T}}$. After any operation $\mathfrak{T}_{\mathrm{cur}}\rightarrow \mathfrak{T}_{\mathrm{new}}$, the vectors $\bm{c}$ and $\bm{w}$ can be locally updated, and the ratio in \eqref{MHratio} can be computed in just $O(1)$ time.

An alternative random walk MH method uses a local proposal density more tuned to the calcium data than is a Gaussian random walk proposal. For each spike time, we construct a distribution from the residual between the data and the current calcium signal, restricted to small time interval $\mathcal{I}$ centered on the current spike time, i.e., $\mathcal{I} = [t_i - L,t_i+L]$ with e.g., $L = 10\Delta$. The proposal distribution can then be expressed as
\[%be
	\log A(t_i\rightarrow t_i') \propto %\exp\left(
	-\frac{1}{2\sigma^2}\sum_{n: n\Delta \in \mathcal{I}}(y(n) - b - c_{\mathfrak{T}_{\backslash i}\cup\{t_i'\}}(n\Delta))^2,%\right),
\]%ee
where $\mathfrak{T}_{\backslash i}$ is the set of spike times excluding $t_i$.
Note that in this case, the proposal distribution is no longer symmetric (since the local intervals are different), and therefore the Hastings ratio is also included in the acceptance probability. However, this can also be computed in $O(1)$ time\protect\footnote{Technically, this computation scales with the resolution at which we want to discretize the proposal distribution. In practice one can use a coarse resolution to choose a bin from the proposal distribution and then sample inside this bin uniformly.} and thus this scheme remains efficient. Finally we note that this proposal distribution can be used to derive a Rao-Blackwellized scheme for updating the continuous time posterior spike distribution, instead of using the actual spike samples. 
This method works quite well empirically.

Lastly, instead of the local random walk methods for moving spikes, we may use Gibbs sampling. We can sample each new location $t_i'$ from the likelihood $p(\bm{y}|\mathfrak{T}_{\backslash i}\cup \{t_i'\},\bm{\theta},\sigma^2)$ using e.g., rejection sampling. While this approach mixes with less samples than using MH (intuitively the spike can move to anywhere, instead of only locally), the computational cost for moving each spike is $O(T)$ which is undesirable.

\subsection{Adding and removing spikes}
The sampler over spike trains also requires transdimensional moves in order to sample over the number of spikes. To add or remove spikes, we follow a standard birth-death MH algorithm \cite{MW03}. We choose a fixed probability $z$ for proposing new spikes and uniform proposal densities for adding and removing spikes. Suppose we want to add a new spike $t_{k+1} = \xi$ to the existing set of spikes $\mathfrak{T}$. This is accepted with probability 
\begin{flalign} %\be
	& \mathbb{P}(\mathfrak{T}\rightarrow \mathfrak{T}\cup\{\xi\}) = \min(1,r_{\mathrm{birth}}), \nonumber \\
%\ee
\text{with}\quad & r_{\mathrm{birth}} =  \frac{p(\mathfrak{T}\cup\{\xi\}|\bm{y},\bm{\theta},\lambda,\sigma^2)(1-z)q_d(\mathfrak{T}\cup\{\xi\},\{\xi\})}{p(\mathfrak{T}|\bm{y},\bm{\theta},\lambda,\sigma^2)z q_b(\mathfrak{T},\{\xi\})}, \nonumber
\end{flalign}
where $q_d(\mathfrak{T}\cup\{\xi\},\{\xi\})$ is the proposal probability for removing $\{\xi\}$ from $\mathfrak{T}\cup\{\xi\}$, and $q_b(\mathfrak{T},\{\xi\})$ is the proposal density for adding $\{\xi\}$ to $\mathfrak{T}$. 
$p(\mathfrak{T}\cup\{\xi\}|\bm{y},\bm{\theta},\lambda,\sigma^2)$ and $p(\mathfrak{T}|\bm{y},\bm{\theta},\lambda,\sigma^2)$ are posterior probabilities of the spike train given the data and the spiking prior respectively after and before the proposed moves.
Under uniform proposals and $|\mathfrak{T}| = K$, we have $q_d(\mathfrak{T}\cup\{\xi\},\{\xi\}) = 1/(K+1)$, $q_b(\mathfrak{T},\{\xi\})=(\Delta T)^{-1}$, giving
\[
	r_{\mathrm{birth}} =  \frac{p(\bm{y}|\mathfrak{T}\cup\{\xi\},\bm{\theta},\sigma^2)(1-z)\lambda\Delta T}{p(\bm{y}|\mathfrak{T},\bm{\theta},\sigma^2)z (K+1)}.
\]%ee
Similarly the acceptance ratio for removing spike $t_i=\eta$ is
\[ 
\begin{split}
	r_{\mathrm{death}} & =  \frac{p(\mathfrak{T}\backslash\{\eta\}|\bm{y},\bm{\theta},\sigma^2)z q_b(\mathfrak{T}\backslash\{\eta\},\{\eta\})}{p(\mathfrak{T}|\bm{y},\bm{\theta},\sigma^2)(1-z) q_d(\mathfrak{T},\{\eta\})} \\
	&  =  \frac{p(\bm{y}|\mathfrak{T}\backslash\{\eta\},\bm{\theta},\sigma^2)z K}{p(\bm{y}|\mathfrak{T},\bm{\theta},\sigma^2)(1-z)\lambda\Delta T}.
\end{split} 
\]
Following \cite{AMM09}, a typical choice of the prior proposal probability is $z = 1/2$, and we repeat the birth-death sampling process 10 times per iteration.  Each iteration of the algorithm is schematically presented in Alg.~\ref{Alg1}. 

\begin{algorithm}
\caption{Schematic representation of each iteration}
\begin{algorithmic}
	\FOR{$i=1$ to current number of spikes $K$}
		\STATE Sample $t_i | \mathfrak{T}_{\backslash i},\lambda,\bm{\theta},\sigma^2$ using MH.
	\ENDFOR
	\FOR{$j=1$ to $10$}
		\STATE Propose addition of spikes. Update $K$.
		\STATE Propose removal of spikes. Update $K$.
	\ENDFOR
	%\STATE Update hyper-parameters $\alpha,\beta$.
	\STATE Sample parameters $\lambda,\bm{\theta},\sigma^2$. 
\end{algorithmic}
\label{Alg1}
\end{algorithm}

Similarly to the case of shifting spikes, the addition and removal can also be implemented in just $O(1)$ time. Using a similar argument it is also easy to show that $\bm{\theta}$ can also be sampled in $O(1)$ time using the method described in section \ref{model}. As a result, the complexity of each iteration scales linearly with the number of spikes and not the number of timebins, rendering this algorithm particularly attractive for recordings that are very sparse and/or have a fine temporal resolution.

%\section{Extensions}
\section{Collapsed Gibbs sampler}
%It is possible to marginalize over the baseline $b$ and initial value $c_1$ without affecting the complexity of the algorithm. 
It is possible to marginalize over the baseline $b$ and initial value $c_1$ and enhance the mixing rates of our algorithm \cite{Liu94}. We present this approach here for the discrete case and note that the continuous case can be treated similarly. For this part we assume $\sigma^2$ to be fixed. By denoting the marginal prior for $[b;c_1] \sim \mathcal{N}(\bm{\mu_b},\Sigma_b)$, %and denoting $\bm{\phi} = [b;c_1] - \bm{\mu_b}$, 
the marginal likelihood is computed as\protect\footnote{Technically this approach is approximate since $[b,c_1]$ have truncated normal priors but here are integrated over the whole $\mathbb{R}^2$. However, in practice the posterior of $[b,c_1]$ puts very little (if not negligible) mass on negative values and the approximation is very tight.}
\[ \begin{split}
	p&(\bm{y}|\bm{s},A,p) = \iint_{b,c_1\in\mathbb{R}^2}p(\bm{y}|\bm{s},p,\bm{\theta})p(b,c_1)db\,dc_1 \\
		%& \propto \exp\left(-\frac{1}{2}\bm{s}^TW\bm{s} + \bm{s}^T\left( \frac{A}{\sigma^2}G^{-T}\bm{y_b}\right)\right) \times \\ 
		%& \quad \times\mathbb{E}\left[\exp\left(-\frac{1}{2\sigma^2}\bm{\phi}^TU\bm{\phi} +\frac{A}{\sigma^2}(\bm{y_m} - G^{-1}\bm{s})^T[\bm{1},\bm{v}]\bm{\phi}\right)\right] \\
		& \propto  \exp\left(-\frac{A^2}{2\sigma^2}\bm{s}^TG^{-T}VG^{-1}\bm{s} +  \frac{A}{\sigma^2}\bm{s}^TG^{-T}V\bm{y_b}\right),
\end{split} \]
%with 
%\be \begin{split}
\begin{flalign}
\text{with}	%&& U & = [\bm{1}_T,\bm{v}]^T [\bm{1}_T,\bm{v}] && \nonumber \\
	&& \bm{y_b} & = \bm{y} - [\bm{1}_T,\bm{v}]\bm{\mu_b}   && \nonumber \\
	&& V &= I+[\bm{1}_T,\bm{v}](\sigma^2\Sigma_b^{-1} +   [\bm{1}_T,\bm{v}]^T [\bm{1}_T,\bm{v}] )^{-1}[\bm{1}_T,\bm{v}]^T  && \nonumber
%	\bm{w} = -\frac{A}{\sigma^2}\left[\begin{array}{c}\bm{1}_T^T \\ \bm{v}^T\end{array}\right]G^{-1}\bm{s}.
\end{flalign}
%\end{split}, \ee
and we used the fact that for $\bm{x}\sim\mathcal{N}(\bm{0},\Phi)$ and a symmetric matrix $C$ such that $\Phi^{-1}-C$ is positive definite, we have
\[
	\mathbb{E}_{\bm{x}}\left[e^{\frac{1}{2}\bm{x}^TC\bm{x} + \bm{b}^T\bm{x}}\right] = |I-\Phi C|^{-1/2}e^{\frac{1}{2}\bm{b}^T(\Phi^{-1}-C)^{-1}\bm{b}}.
\]%ee
The marginalized likelihood has the same functional form and therefore $\bm{s}$ and $A$ can again be sampled with the same methods. Moreover, by exploiting the structure of $V$ it is easy to see that each multiplication of the form $VG^{-1}\bm{s}$ can be performed in $O(1/\log(1/\gamma))$ time and therefore the complexity of each full Gibbs sweep still scales as $O(T)$ and the algorithm remains efficient. After the initial burn-in period, the (Rao-Blackwellized) posterior of the marginalized parameters $b,c_1$ can be approximated as
\[
	p(b,c_1| \bm{s},A) \approx \frac{1}{M}\sum_{i=1}^Mp(b,c_1|\bm{s^{(i)}},A^{(i)}),
\] 
where $\bm{s^{(i)}},A^{(i)}$ are the sampled values and each $p(b,c_1|\bm{s^{(i)}},A^{(i)})$ can be computed by conditioning $p(\bm{\theta}|\bm{s^{(i)}})$. %In practice this marginalization approach (collapsed Gibbs sampler) leads to significantly faster mixing rates compared to the standard Gibbs methods. 
Note that $\sigma^2$ is treated as a known parameter here. If we assume an inverse Gamma prior then the posterior is no longer inverse Gamma. However $\sigma^2$ can still be sampled with standard rejection sampling methods. Finally marginalization over $A$ is also possible, however the marginalized posterior of $\bm{s}$ has no longer the nice quadratic form and the computational complexity of the algorithm increases. %A similar approach can also be derived for the continuous time sampler.

\section{Results}
\begin{figure}[!t]
	\includegraphics[trim = 30mm 22mm 20mm 9mm, clip, width=.5\textwidth]{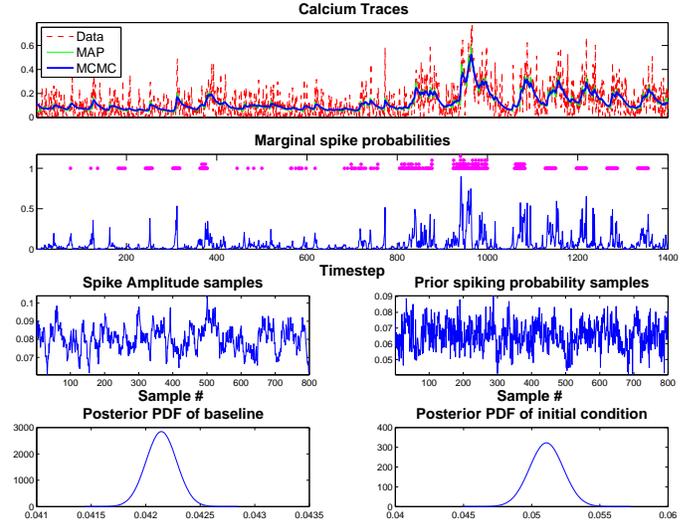}
	\caption{Application of the collapsed, discrete Metropolized-Gibbs sampler on real spinal cord data (in-vitro). Top row: Real data (red dashed), MAP estimate (green) and mean from 800 Gibbs samples (blue). Second row: Marginal spike probabilities per timestep. The true spike times are also shown (purple circles). Third row: 800 samples from the spike amplitude (left) and the prior spiking probability (right). Bottom row: Rao-Blackwellized estimates of the posterior distributions for the baseline (left) and initial value (right). The collapsed MG sampler mixes fast and provides low variance estimates for the model parameters.}
	\label{fig1}
\end{figure}

We apply our methods to calcium imaging data from spinal cord neurons in-vitro, collected using the calcium indicator GCaMP6s with a temporal resolution 15Hz \cite{MPJ13}. The neurons were stimulated antidromically (so we know the spike times) and fired small bursts of spikes. In several timebins multiple spikes were fired. Fig.~\ref{fig1} shows an application of the discrete algorithm with $b,c_1$ marginalized out. 1000 samples were collected with the first 200 discarded (burn-in period). This particular trace is of low SNR, however the algorithm predicts mosts of the spikes and provides low variance estimates of the model parameters. 

\begin{figure}[!t]
	\includegraphics[trim = 30mm 9mm 20mm 9mm, clip, width=.5\textwidth]{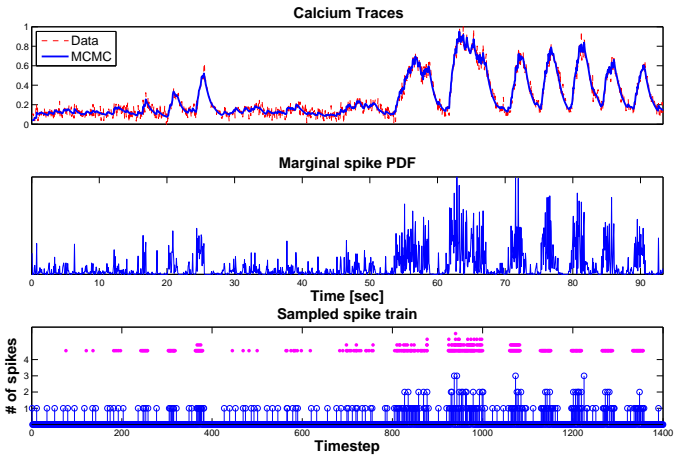}
	\caption{Application of the collapsed, discrete Metropolized-Gibbs sampler on real spinal cord data (in-vitro). Top row: Real data (red dashed) and mean from 700 Gibbs samples (blue). Second row: Estimated marginal spike PDF in continuous time. Third row: A typical samples of the spike train in continuous time, binned in the resolution defined by the imaging rate.The true spike times are also shown (purple circles). The continuous time sampler can assign multiple spikes at each timebin and provide better approximation of the spiking behavior in high-SNR conditions.}
	\label{fig2}
\end{figure}

A limitation of the discrete algorithm is that it can only produce one spike per timebin. To resolve this issue we run the continuous time sampler to the same dataset. All the traces from imaged pixels that correspond to the same neuron were averaged to produce a high-SNR trace. The algorithm produced 500 samples after a burn in period of 200 samples. The results are shown in Fig.~\ref{fig2}. Again the algorithm predicts well the produced calcium trace and provides estimates of the spike posterior in continuous time. By binning the produced sample in the original bin size (bottom row) we see that the algorithm can assign multiple spikes per timebin and better approximate the true spikes.

\section{Conclusions - Future Work}
We presented two classes of Bayesian methods for spike train inference from calcium imaging data. Our methods provide a principled approach for estimating the posterior distribution of the spike trains and provide robust estimates of the model parameters, especially in low-SNR conditions. We also derived collapsed Gibbs samplers that exhibit faster mixing with no significant computational cost per sample. 

In future work, we plan to explore the use of Hamiltonian Monte Carlo \cite{Ari13} and particle Markov chain Monte Carlo \cite{ADH10} methods for more efficient spike sampling, and extend our methods to allow for a slowly time-varying baseline, a phenomenon that is often observed in vivo experimental conditions. We also plan to scale up to a spatial setup where each measurement corresponds to a pixel that is part of a neuron (or is shared across a small number of neurons) \cite{PMG13}, and in the case of compressive calcium imaging \cite{PP13a}, where each measurement is formed by projecting the calcium activity of the whole imaged spatial onto a random vector.

%Our  and can also be used in cases where the calcium measurements are intermittent \cite{KPP13} or are available in the form of random projections \cite{PP13a}.

\section*{Acknowledgment}
We thank S. Chandramouli and D. Carlson for useful discussions, and T. Machado for sharing his spinal cord data. LP is supported from an NSF career award. This work is also supported by ARO MURI W911NF-12-1-0594.

{\small
\bibliographystyle{IEEEtran}
\bibliography{../mybib.bib}
}

\end{document}